\documentclass[12pt,preprint]{aastex}

\shorttitle{Characteristic Scales in Stellar Clustering}
\shortauthors{Odekon}

\begin{document}


\title{Characteristic Scales in Stellar Clustering: A Transition Near the Disk Scale Height} 


\author{Mary Crone Odekon}
\affil{Department of Physics, Skidmore College, Saratoga Springs, NY 12866}
\email{mcrone@skidmore.edu}



\begin{abstract}
The autocorrelation function provides an objective test for the existence of special scales in the hierarchical clustering of young stars.  We apply this measure to single-star photometry for the brightest main sequence stars in the Small Magellanic Cloud (SMC), the Large Magellanic Cloud (LMC), M33, and M31, using data from the Magellanic Clouds Photometric Survey and the Massey Local Group Survey.  Our primary result is the identification of a transition to a higher correlation dimension (weaker clustering) at one kpc in the LMC and M31, and at 300 pc in M33.  We suggest that this transition marks the large-scale regime where disk geometry and dynamics set the scale for structure.  On smaller scales, the correlation functions for each galaxy are scale-free over at least two orders of magnitude, with a projected correlation dimension varying from 1.0 for M31 to 1.8 for the SMC.  This variation is probably caused by a combination of differences in stellar ages and masses, physical environment, and extinction. 
\end{abstract}



\section{Introduction}

While it is clear that young stars are clustered hierarchically, 
with OB associations and clusters inside complexes and superassociations,
the extent to which there are universal clustering scales is controversial.  Historically,  typical sizes assigned to stellar groupings have depended at least partially on selection criteria and image quality.  Efremov (1995) provides a summary of this history, and concludes that it make sense to consider two distinct scales within the hierarchy:  elementary cells on 
scales of 10-100 pc that contain stars with ages up to 10 Myr, and large complexes on scales of several hundred parsecs that 
contain stars with ages up to about 100 Myr. The identification of clusters on different scales is an area of active research, using a variety of surveys and instruments.  For example, Livanou et al. (2007a, 2007b) combine optical and infrared obserations to find clusters in the Magellanic Clouds on scales ranging from 150 pc to 1500 pc, that they classify as aggregates, complexes, or supercomplexes based on size.   In fact, many studies support an approximately scale-free clustering pattern for both the gas and stars in galaxies, a behavior typically attributed 
to the turbulent cascade of energy within the gas (see Elmegreen \& Scalo 2004 for a review).  

A quantitative description of stellar clustering --- including the identification of any characteristic scales --- is a key part of several current problems in the formation and evolution of stars, including how closely young stars follow the distribution of gas and dust,  how quickly they disperse, and under what conditions they remain gravitationally bound (e.g. Allen et al. 2007, Shu et al. 2007, Yang et al. 2007, Mu\~{n}oz et al. 2007, Pellerin et al. 2007, Carlson et al. 2007).  More generally, it addresses the circumstances under which we should   describe young stars in terms of clusters at all.  

One way to quantify clustering is through multi-scale variants of cluster identification methods (e.g. Battinelli \& Efremov 1999, Elmegreen et al. 2006). For example, 
Bastian et al. (2007) studied the clustering pattern of OB stars in M33 using a multi-scale implementation of the Minimum Spanning Tree method, and found no characteristic scales. 

Power spectra and correlation functions, on the other hand --- while unsuited for identifying \textit{individual} clusters --- are more directly geared to searching for special scales, and have been used for this purpose on the integrated light of galaxies (e.g. Elmegreen et al. 2003a; Elmegreen et al. 2003b; Willet et al. 2005).  Two challenges in interpreting these studies are the fact that integrated starlight includes stars that are older and less clustered, and that individual bright stars create a form of contamination to the power spectrum that can mimic a transition in clustering strength (Elmegreen et al. 2003a).  Odekon (2006) and Parodi \& Binggeli (2003) used the integrated autocorrelation function to quantify the clustering strength of individual young stars and individual blue clumps of stars, respectively, in dwarf galaxies, but the number of objects was too small (32-125 young stars per galaxy, and up to 372 blue clumps per galaxy) to provide a statistically powerful test for changes in clustering strength with scale. 

The Magellanic Clouds Photometric Survey (MCPS; Zaritsky et al. 1997) and Massey Local Group Survey (Massey et al. 2006; Massey et al. 2007) provide public catalogs of deep single-star photometry with coverage across the entire star-forming region of most Local Group galaxies.  Single-star photometry allows the separation of clustered, young stars from dynamically relaxed, older stars.  It also prevents the problem of a dominant signal from individual bright stars, because each star counts equally.  Four galaxies (the SMC, the LMC, M33, and M31) contain a large number of bright, young stars, enough to provide a statistically powerful test of clustering strength over a wide range of scales. 
Indeed, Harris \& Zaritsky (1999) calculated the spatial correlation function for stars in a subsection of the LMC, and found that it is consistent with an exponential function that changes systematically with the mean age of the stellar populations; however, they were limited to a region 1/18 of the final survey area. 

\section{The Data}

Table 1 summarizes the data.  The MCPS catalogs give \textit{UBVI} stellar photometry for the SMC and the LMC down to an apparent magnitude of 21 (an absolute magnitude of $2.0-2.5$ at the distances of the Magellanic Clouds).  They include over five million stars for the SMC and over 24 million stars for the LMC.  The \textit{UBVRI} Massey catalogs reach an apparent magnitude of 23 (an absolute magnitude of about $-1.5$ at the distance of M31 and M33) with photometric errors less than 10\%, and includes 371,781 stars for M31 and 146,622 stars for M33.  The Massey catalogs include the Magellanic Clouds, but their coverage is not as complete as the MCPS.

We select young stars with $V-I$ colors between $-0.3$ and 0.0 and absolute magnitudes between $-6$ and $-4$; this region of the color-magnitude diagram---the tip of the main sequence---is small enough to minimize foreground contamination (Massey et al. 2006; Zaritsky et al. 2002), and includes stars younger than about 30 Myr (Girardi et al. 2002).  We also tried selecting color by $B-V$ and the reddening-free parameter $Q$, without a significant change in our results for the correlation function.  

By eye, the spatial distributions of the bright main sequence stars (Figure 1) exhibit clustering over a wide variety of scales.  The data for M31, a large and highly-inclined galaxy, are probably the most affected by extinction and crowding.   


Both surveys use multiple scans to cover the star-forming region of each galaxy, and are subject to variations in image quality from scan to scan.  Photometric inhomogeneities at scan edges, as well as incompleteness in regions of high crowding and extinction, appear in maps of the locations of fainter stars (e.g. the red clump stars in Figure 5 of Zaritsky et al. 2004).  In order to assess their effect on features in the correlation function of young stars, we also calculated the correlation function for red giants, which are slightly fainter than the bright the main sequence stars, and should exhibit a smooth, relaxed distribution.  As described below, this exercise shows that the correlation functions of M33 and M31 are indeed noticeably affected by scan edges, but only on scales larger than the features we discuss here.  

\section{The Autocorrelation Function}

To measure clustering as a function of scale we use the angular two-point autocorrrelation function $c(\theta)/\theta$, where $c(\theta)d\theta$ is the average number of stars with angular distances 
between $\theta$ and $\theta+d\theta$ from any particular star. The correlation integral $C(\theta)$ and correlation dimension $D$ are closely related to this function:  $C(\theta) \propto \theta^D$, where $C(\theta) \propto \int_0^\theta c(\theta')d\theta'$.  For a scale-free distribution, both $c(\theta)/\theta$ and $C(\theta)$ have the form of a power law, where the power of $C(\theta)$ is greater by two.  For example, the number of points within a distance $r$ on a flat, two-dimensional surface covered by a random (Poisson) distribution of points scales as $r^2$, yielding a power of two for the correlation integral $C(r)$ and a power of zero for $c(r)/r$.   For a distribution that is more highly clustered, the power law exponent is lower.  For a distribution that not perfectly scale-free, the power law changes with scale.  If the goal is to search for such changes, it is preferable to use the derivative form, as we do, rather than the less-sensitive correlation integral.    

To test the ability of this function to measure clustering strength and find preferred scales, we calculated it for several finite distributions of points (Figure 2).   On small scales, the expected power (the slope on these log-log plots) is indeed recovered: zero for Poisson distributions and $-0.415$ ($=2-1.585$) for the Sierpinski Triangle, a well-known fractal of dimension 1.585.  On the largest scales, the finite size of the distribution creates a smooth cutoff.  For models with clumps and rings, the correlation function picks out the sizes of these features and the distances between them.

The area of an annulus with radius $r$ is not exactly proportional to $r$ because of its nonzero width $dr$.  Specifically, it scales as $(r+dr)^2-r^2$, producing a power of $r/(r+dr)$ instead of one.  In our analysis, we use an increment of 0\fdg001 (0\fdg0001 for M33, which has more resolved stars in the sample), so the discrepancy between one and $r/(r+dr)$ is less than 10\% on scales larger than 0\fdg01 (0\fdg001 for M33).  A still smaller correction comes from the fact that angular distances are projected onto the sphere of the sky, so that the circumference scales as $\sin\theta$ instead of $\theta$;  for scales smaller than 5\degr\, this produces a difference in the slope of less than 1\%

\section{Results}

Figure 3 shows the correlation functions for the brightest main sequence stars in each galaxy and, for comparison, the bright red giants.  The correlation functions for the red giants are similar to those for the Poisson disk and ellipsoid --- flat at small scales with a smooth cutoff at large scales.  An exception is the slight decrease at about a tenth of a degree in M31 and M32, reflecting the inhomogeneities at scan edges described in section 2. 

The correlation function for young stars is different from the red giants in two important ways:  its slope is lower, indicating a more highly clustered distribution, and it exhibits features at about one kpc for the LMC and M31 and at about 300 pc for M33.  For the LMC, a slight decrease in slope precedes a brief flattening, a behavior qualitatively similar to that for clumps and rings.  For M33 and M31, only a transition to a flatter slope is evident. On scales below the transition, the power-law exponents for the young stars are -0.2, -0.4, -0.6, and -1.0 (that is, correlation dimensions of 1.8, 1.6, 1.4, and 1.0) for the SMC, LMC, M33, and M31, respectively.   

Splitting the stars into sub-populations by location on the sky produces correlation functions qualitatively similar to those for the full populations --- in particular, the existence or non-existence of the transition features remain unchanged.   The largest difference is for the LMC, where the eastern half produces a peak at the approximate radius of the two large clumps on that side (see Figure 1), but both sides still show a feature at about one kpc.  

Splitting the stars into sub-populations by color or magnitude produces steeper power laws for bluer and brighter stars.  The largest difference can be dectected in M33, where the power is as steep as $-1.0$ for main sequence stars with $-6<M_I<-5$ and $-0.3<V-I<-0.1$ but only $-0.5$ for main sequence stars with $-4.5<M_I<-4.0$ and $-0.3<V-I<0.0$ (Figure 4).  The primary cause for this trend is probably a younger average age for the brighter and bluer stars.  Indeed, many types of measurements have suggested a gradual decrease in clustering with age, presumably illustrating the dissipation of clusters with time (e.g. Harris \& Zaritsky 1999, Zhang et al. 2001, Schmeja \& Klessen 2006, Odekon 2006, de la Fuente Marcos \& de la Fuente Marcos 2006, S\'{a}nchez et al. 2007).  This trend could also reflect mass segregation, with the bluer and brighter stars more clustered because of a higher average mass rather than a younger average age.  In addition, the brighter stars may have more complete photometry that allows us to see the more intense clustering on smaller scales, although it seems unlikely that this would make a difference on scales as large as hundreds of parsecs.  

We tried selecting stars of an intermediate age in each galaxy, considerably further down the main sequence at $-2.5<I<-2.0$.  For the Magellanic Clouds, the results were similar to those for the brighter main sequence stars, but slightly less clustered.  For M33 and M31, this population is significantly polluted with randomly-scattered foreground stars, and is affected by the inhomogeneities at scan edges.  Figure 4 includes this population for M33.  

For both the LMC and the SMC, we also computed the correlation function using data from the Massey catalogs, which have different spatial coverage;  these gave the same results as the MCPS.  

\section{Discussion}

The correlation function for all four galaxies is scale-free over about two orders of magnitude, starting from the smallest scales we can resolve with these data: about 20 pc in M31, and just a few parsecs in the other galaxies.  A transition feature in the correlation function appears at one kpc in the LMC and M31 and at 300 pc in M33, above which the slope flattens.  This scale roughly corresponds to the fundamental complexes of Efremov (1995) and the size Elmegreen (2006) called the largest globular regions of star formation in galaxies. 

The transition in the LMC occurs at a considerable fraction of the galaxy size --- about 16\%.   Only a small number of clumps at this size or larger could exist in the galaxy, and their properties drive the appearance of the transition.  For M33, on the other hand, the transition scale is small enough relative to the galaxy --- about 3\% --- that the absence of features on larger scales is statistically meaningful.  

The transition features are not caused by simple edge or boundary effects.  In particular, they do not represent the geometry of the galaxies as projected on the sky.  For a projected distribution with a decreasing density profile (e.g. the first two models of Figure 1), the slope steepens.  Similarly, the scan inhomogeneities apparent in the red giant correlation functions for M33 and M31 create a steepening slope, a mini-edge.  The transition features in the data, on the other hand, are an abrupt \textit{flattening} on large scales.   

We suggest a different connection between the transition scale and disk geometry.  At the transition scale, density fluctuation modes become highly dependent on disk structure.  From a dynamical standpoint, the modes that develop within a rotating disk depend partly on the disk thickness (e.g. Tasker \& Bryan 2008).  And of course, spherical clusters (Elmegreen's ``globular regions'') that are too large cannot exist inside the two-dimensional disk.  

Padoan et al. (2001) and Elmegreen et al. (2001) found that the clustering of HI in the LMC shows a feature at about 180 pc that they identify as the transition from small-scale three-dimensional turbulence to large-scale two-dimensional disturbances in the disk.  More recently, Kim \& Park (2007) found a transition in HI for the LMC at 290 pc.  We may be seeing the analogous transition for stars. 

Following Efremov (1995), we might expect a smaller characteristic scale for the youngest stars and a larger one for slightly older stars.  However, when we subdivide the young stars by magnitude or color, we see that the location of the transition changes only slightly, if at all, even when there is a considerable change in slope.  Figure 4 shows the most extreme case, where the bluest and brightest subpopulation of main sequence stars in M33 have a considerably steeper slope, but no characteristic scales smaller than the transtion scale at several hundred parsecs.  It is true that the correlation function for M31 does not reach scales even as small as 15 pc.  A combination of photometry from multiple \textit{Hubble Space Telescope} pointings (e.g. those used by Williams \& Hodge 2001 to identify clusters in M31 on a scale of about five pc) could provide a statistically meaningful test on this scale. 

To address this question further, it would be useful to select stars with a greater variety of age categories, including very young populations as well as intermediate-aged populations that are not highly contaminated by foreground stars.  Possible approaches to this include infrared identification of young stellar objects like those in Bolatto et al. (2007), spectroscopic identification, and selection of Cepheids with a specific period.  Another improvement would be to use results from population synthesis modeling of an entire color-magnitude diagram to restrict the ages that are statistically dominant in a certain regions of the bright main sequence.  

A related question, one that could also be addressed using a broader range of scales and age categories, is the statistical detection of cluster dissipation with age, an effect that might cause an additional transition, at a scale that increases with age, to \textit{weaker} clustering on small scales.  Mass segregation would probably have a very similar effect.  In both the LMC and M33 we see a hint of this behavior in the correlation dimension for the fainter subset of main sequence stars, but the correlation function is statistically messy on these scales, with this smaller sub-populations of stars. 

On scales below the transition, the correlation dimensions for young massive stars varies from about 1.0 for M31 and the brightest subpopulation of M33 to 1.8 for the SMC.  The variation in clustering strength may reflect differences in the dominant ages or masses of stars, but a quantitative understanding of the correlation dimension really requires a detailed assessment of crowding and extinction, beyond the scope of this paper.  Crowding and extinction are likely to cause increasing incompleteness on smaller scales, even for stellar populations that are completely recovered in simple artificial-star tests (see the discussion, for example, in Odekon 2006).  This is likely to be an especially large problem for M31, which is highly inclined as well as large and dusty.  On the other hand, as a rough completeness check in the SMC, we compared the MCPS photometry in the cluster NGC 346 with the high-resolution Advanced Camera for Surveys photometry of that region (Gouliermis et al. 2006).  They show excellent agreement, the only missing stars residing in a tiny cluster about one pc across.    

These correlation dimensions are based on two-dimensional, projected distributions.  To estimate the true three-dimensional value, it is necessary to consider the extent to which the observations represent projections rather than slices --- or outer skins --- of each galaxy.  While a slice through a fractal gives a dimension less by one, a projection of a fractal approximately reproduces the three-dimensional value if it is less than two, and gives two otherwise (see S\'{a}nchez et al. 2005 for detailed models).  For this reason, galaxies that are more transparent may appear to have a larger two-dimensional correlation dimension, simply because we see their distribution more as a projection than as a thin, outer skin.  This is probably not, however, the only reason for the higher dimension we observe in the SMC compared with M31, given the systematically lower correlation dimensions for younger subsets of stars, as expected for populations relaxing with age. 

Finally, our results show that it is not surprising to see a fairly wide range of correlation dimensions from galaxy to galaxy, (e.g. those in Parodi \& Binggeli 2003 and Odekon 2006), given the existence of a transition scale in some galaxies and the differences in clustering strength for different population of bright main sequence stars.  

\acknowledgments

This project relies on data products from the Magellanic Clouds Photometric Survey of Zaritsky et al. (2002, 2004), accessible at http://ngala.as.arizona.edu/dennis/mcsurvey.html, and from the Local Group Survey of Massey et al. (2006, 2007), accessible at \\
http://www.lowell.edu/users/massey/lgsurvey.html.
 


\clearpage
\begin{table}
\begin{center}
\caption{Galaxy Data\label{tbl-1}}
\begin{tabular}{clrccl}
\tableline\tableline
Galaxy&Type	& $M_V$		&Distance modulus	&Number of	&Reference \\
			&			&					&$m-M$							&selected stars 	&					\\
\tableline
SMC 	&Ir IV-V 		&-17.1	&19.0	&798	&Zaritsky et al. 2002\\
LMC 	&Ir III-IV	&-18.5	&18.5	&2332	&Zaritsky et al. 2004\\
M33 	&Sb I-II 		&-18.9	&24.7	&3796 &Massey et al. 2006\\
M31 	&Sc II-III	&-21.2	&24.5 &1188	&Massey et al. 2006\\
\tableline
\end{tabular}
\tablecomments{Galaxy properties are from Massey et al. 2006.} 
\end{center}
\end{table}
\clearpage


\begin{figure}
\epsscale{0.9}
\plotone{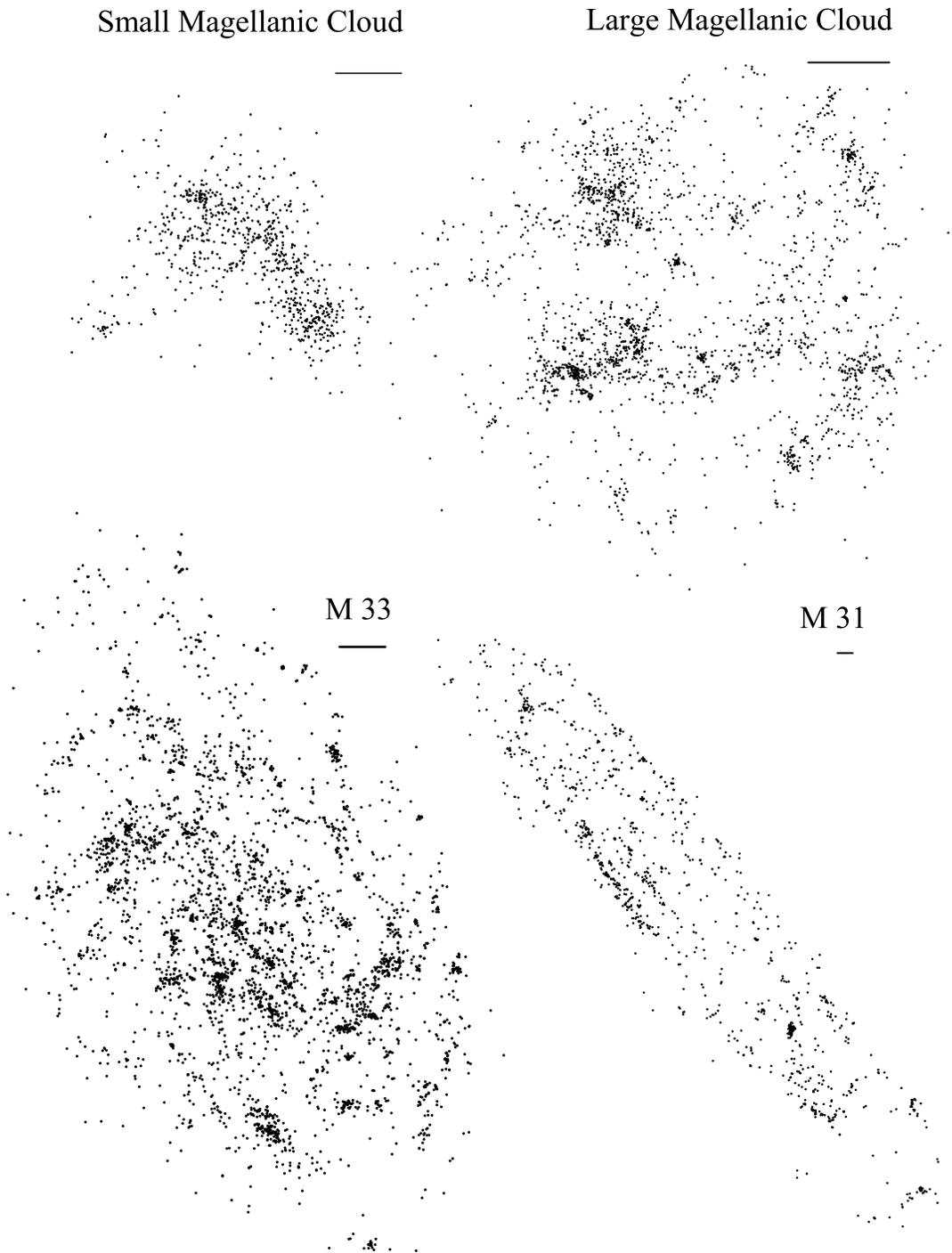}
\caption{The spatial distributions of bright main sequence stars.  Horizontal bars give a scale of one kpc for each galaxy.  North is up and East is to the left.  Clustering is evident on a wide variety of scales.}
\end{figure}

\begin{figure}
\epsscale{1.0}
\plotone{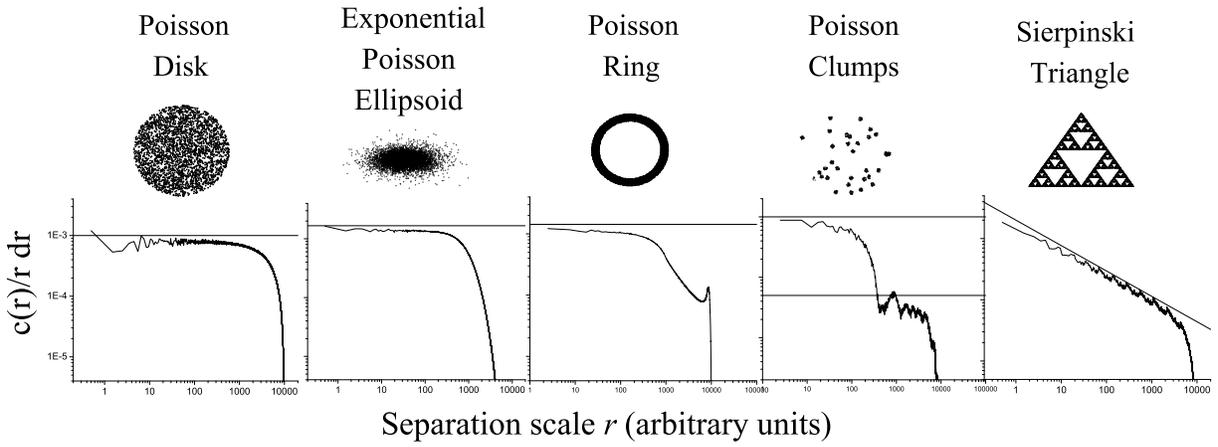}
\caption{Correlation functions for several finite point distributions.  Straight lines indicate the expected behavior for an infinite, two dimensional Poisson distribution (horizontal lines) and an infinite Sierpinski triangle of fractal dimension 1.585 (slanting line).}
\end{figure}

\begin{figure}
\plotone{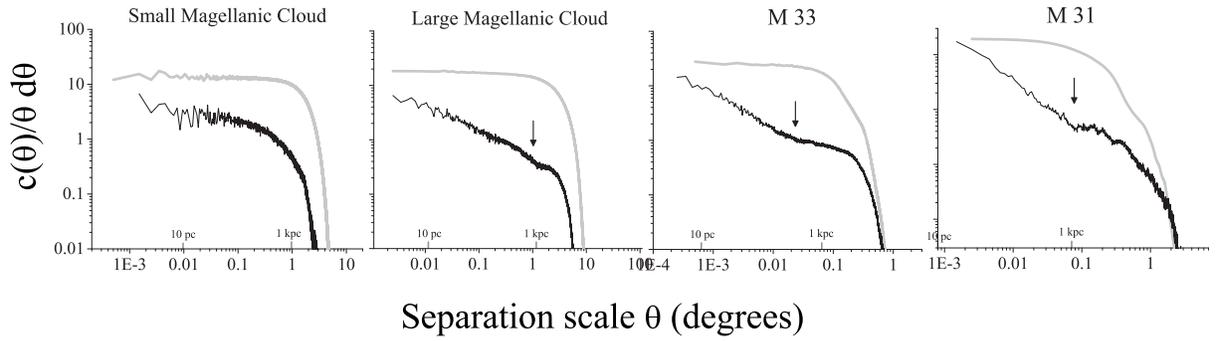}
\caption{Correlation functions for bright main sequence stars (black curves) and, for comparison, red giants (grey curves).  Physical scales of ten pc and one kpc at the distance of each galaxy are labeled, and arrows indicate the transition features. 
} 
\end{figure}

\begin{figure}
\plotone{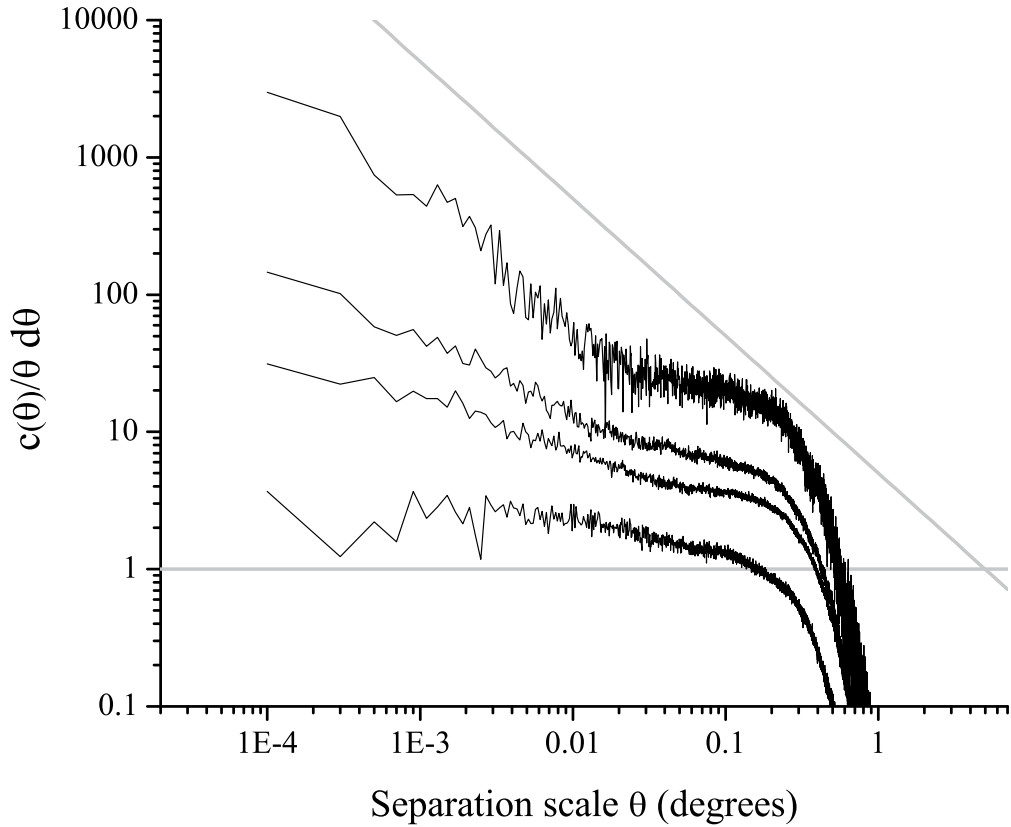}
\caption{Correlation functions for different sections of the main sequence in M33.  The top two curves show subsections of the bright main sequence population of Figure 3, while the bottom two show fainter sections.  From top to bottom, the ranges for $M_I$ are -6.0 to -5.0, -4.5 to -5.0, -4.0 to -3.5, and -2.5 to -2.0.  Colors are $-0.3<V-I<0.0$, except for the brightest subsection, which has the more restrictive red cutoff $V-I<-0.1$.  Straight gray lines show power laws of -1 and 0 (correlation dimensions of 1 and 2).  The correlation functions have been shifted vertically to show all four without overlap.  
} 
\end{figure}
\end{document}